\documentstyle[12pt,axodraw,epsfig]{article}
\newcommand{\eqref}[1]{(\ref{#1})}

\def\ltap{\raisebox{-.55ex}{\rlap{$\sim$}} \raisebox{.4ex}{$<$}}
\def\gtap{\raisebox{-.55ex}{\rlap{$\sim$}} \raisebox{.4ex}{$>$}}
\def\gsim{\mathrel{\gtap}}
\def\lsim{\mathrel{\ltap}}

\def\const{\mbox{const}}

\def\half{{1 \over 2}}

\begin{document}

\title{Cosmology with non-minimal scalar field: graceful entrance into
inflation.}

\author{
  M.V.~Libanov, V.A.~Rubakov and  P.G.~Tinyakov \\
  {\small {\em Institute for Nuclear Research of the Russian 
   Academy of Sciences,}}\\
  {\small {\em 60th October Anniversary prospect 
   7a, Moscow 117312, Russia.}}
  }
\date{}
\maketitle

\begin{abstract}
We propose a scenario of the beginning of inflation in which the
non-vacuum value of the scalar field that drives inflation develops
dynamically due to the non-minimal coupling to gravity. In this
scenario, inflation emerges as an intermediate stage of the evolution
of the Universe, well after the Planck epoch, from a fairly general
initial state.
\end{abstract}

One of the intriguing aspects of inflation is its beginning. When did
inflation start? Why was the inflaton, the scalar field
responsible for inflation, essentially homogeneous over the initial
Hubble volume? Why was the inflaton field originally placed far away
from its present value, the physical minimum of its effective
potential?  Designing various ways to address these issues is not only
of academic interest: different scenarios of the beginning of
inflation may lead to different observable consequences, such as
properties of the spectrum of density perturbations and/or
gravitational waves, deviation of $\Omega_{\rm present}$ from 1, etc.

In a currently popular set of scenarios (for reviews and references
see, e.g., ref.\cite{reviews}), inflation began immediately after the
epoch governed by quantum gravity. In that case the initial conditions
for the classical evolution are in principle determined by the Planck
scale physics or semiclassical quantum gravity phenomena. While it is
feasible that quantum fluctuations of gravitational and matter fields
eventually give rise to the classical inflationary expansion, their
quantitative analysis faces problems related to the lack of the
detailed theory of quantum gravity.

Another possibility is that inflation is an intermediate stage of the
evolution of the Universe, i.e., that it started well after the Planck
epoch. This possibility is realized in the models of ``old''~\cite{Guth}
and ``new''~\cite{new} inflation
 and their descendants, where
the inflaton field is driven to a ``false'', metastable minimum of its
effective potential, e.g., by thermal effects. These models, however,
are in potential conflict with flatness of the inflaton potential that
is required for generating acceptably small density perturbations.

In this paper we suggest another mechanism for the ``intermediate
stage'' inflation.  We show that in a class of models with the
inflaton field non-minimally coupled to gravity, inflationary
expansion emerges automatically well after the Planck epoch from a
fairly general initial state. In particular, the classical scalar
field may initially be placed at the true minimum of its
potential. What is required is the pre-inflationary Friedmann-like
stage with deviation from radiation domination. The idea is that the
non-minimal gravitational coupling of the inflaton field drives it out
to non-zero value, so that the Friedmann expansion is followed by the
inflationary stage.  The mechanism works provided that the scalar
potential is sufficiently flat and its functional form is related in a
certain way to the field-dependent Planck mass describing the
non-minimal coupling of the scalar field to gravity.

In our scenario, the inflationary stage itself is similar to chaotic
inflation~\cite{Linde-chaotic}, as the scalar field slowly rolls
down the monotonous effective potential. The difference to many models of
chaotic inflation is that in our case inflation begins when the
value of the inflaton potential is well below the Planck energy
density. Hence, the number of $e$-foldings is not extraordinarily large.
This may lead to
$\Omega_{\rm present}\neq 1$ and/or features in the observable spectrum of
density perturbations, in addition to another observational
consequence \cite{Kaiser} of the non-minimally coupled
inflaton --- deviation of the spectral index from 1.

The models of the class that we consider in this paper
contain one real scalar field $\phi$ and are described by the action
\[
S = \int d^4x \sqrt{-g} \left\{ - {M_{Pl}^2\over 16\pi} A(\phi) \cdot
R + \half \partial_{\mu}\phi
\partial^{\mu}\phi - V(\phi) \right\}. 
\]
We will also add matter that produces non-vanishing scalar curvature
$R$ at the initial Friedmann stage.
In what follows we will use the units in which
$
3M_{Pl}^2/ 8\pi =1.
$

As we do not specify the particle physics origin of the inflaton field,
the functions $A(\phi)$ and
$V(\phi)$ are essentially arbitrary. We assume that the
scalar potential has one minimum at $\phi=0$ and the cosmological
constant vanishes, 
\[
V(0)=0\;. 
\]
Then the function $A(\phi)$ behaves as follows near $\phi=0$, 
\begin{equation}
A(\phi) = 1 + \delta \phi + \half \xi\phi^2 +\ldots
\label{A=1+df}
\end{equation}
The most favorable case for generating inflation from a state with
$\phi\simeq 0$
is $\delta\neq 0$. It is this case that we consider in  this
paper; we define the field in such a way that 
\begin{equation}
\delta>0\;.
\label{delta>0} 
\end{equation}
To be specific, we assume that $A(\phi)$ increases monotonously at
$\phi>0$ and does not contain large or small parameters; in
particular, $\delta\sim 1$ in Planck units. On the other hand, the
scalar potential $V(\phi)$ is required to contain small parameters so
that $V(\phi)\ll1$ at $\phi\lsim 1$.  Furthermore, we choose the
functional forms of $V(\phi)$ and $A(\phi)$ in such a way that they
are correlated above a certain sub-Planckian scale $\mu$,
\begin{equation}
V(\phi) \approx \const \cdot A^2(\phi)\ \ \ \mbox{at}\ \ \phi\gg\mu\;, 
\label{V=A2}
\end{equation}
We will see that successful inflation occurs as an intermediate stage
for $\mu\lsim10^{-3}$ and that it is sufficient that the
relation~(\ref{V=A2}) holds well below the Planck scale only. In other
words, we need not specify any relation between $V(\phi)$ and
$A(\phi)$ at $\phi\sim1$.

A concrete example which we will occasionally  refer to  is
\begin{eqnarray}
A(\phi)&=&\left(1+\frac{\delta}{2}\phi\right)\;,\nonumber\\
V(\phi)&=&\lambda\frac{\phi^2}{\mu^2+\phi^2}
\left(1+\frac{\delta}{2}\phi\right)^2\;,
\label{V1**}
\end{eqnarray}
where $\delta\sim 1,\ \ \lambda\ll1$ and $\mu\ll1$. 
We stress 
that the concrete form~(\ref{V1**}) will be used for illustrative purposes 
only, while our analysis and results apply to the general case specified
by eqs.~(\ref{V=A2}) and (\ref{delta>0}).

Equation (\ref{V=A2}) is easy to understand in the
Einstein frame (i.e., after performing the conformal transformation
$g_{\mu\nu}\to A^{-1}g_{\mu\nu}$): the effective potential relevant to
this frame, 
\begin{equation}
U(\phi) = {V(\phi)\over A^2(\phi)}\;,
\label{effective}
\end{equation}
has the form reminiscent of the potential along an approximate flat direction,
 as shown
in Fig.1. (We will not use the Einstein frame in what follows, because the
description of matter in that frame is inconvenient.)

\iftrue
\begin{figure}[ht]
\begin{center}
\epsfig{file=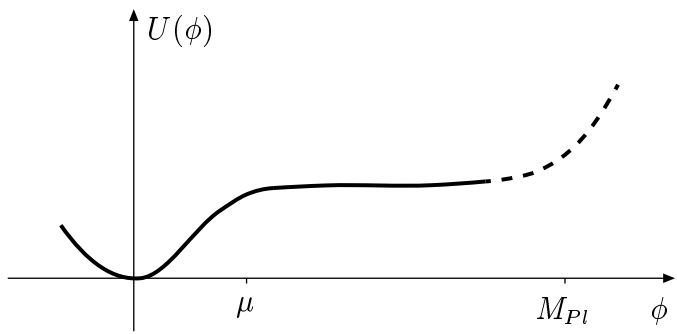,%
bbllx=220pt,bblly=585pt,%
bburx=485pt,bbury=715pt,
clip=}
\end{center}
\caption{Einstein-frame effective potential.}
\end{figure}
\fi

To begin with, let us consider the open FRW Universe which is initially filled 
with matter, and study homogeneous scalar field $\phi$. We will argue later on
that the homogeneity and/or isotropy of the initial state of the Universe are 
in fact irrelevant. On the other hand, the assumption of open Universe 
{\it is} essential: while our mechanism works even better in the spatially flat
case, the closed Universe of sufficiently small size shrinks down to 
singularity before the conditions for the beginning of inflation set up.

The effect of matter is to produce the scalar curvature at the early
stage of the expansion. As an example, we consider the equation of
state $p=\gamma\epsilon$ with $0\leq\gamma<1/3$. Then the field
equations take the following form
\begin{equation}
\left( H^2 + {\kappa\over a^2}\right) A(\phi) + H A'(\phi) \dot\phi =
{\rho_0\over a^{\alpha}}+\frac{1}{2}\dot{\phi}^2+V(\phi)\;,
\label{Einstein-a}
\end{equation}
\begin{equation}
\ddot\phi + 3H\dot\phi - \left( \dot H +2H^2 +{\kappa\over a^2}\right)
A'(\phi) = -V'(\phi),
\label{Einstein-f}
\end{equation}
where $a$ is the scale factor, $H=\dot a/a$ is the Hubble parameter, 
$\alpha = 3(1+\gamma)$,
and $\rho_0$ and $\kappa=-1$ characterize the initial density of matter
and spatial geometry, respectively. 
We consider the evolution after the Planck epoch, i.e.,
set $a(t_i)=1$ at the  initial time $t_i$. As an example we impose the
initial condition 
\begin{equation}
\phi(t_i) = 0, ~~~~~~\dot\phi(t_i)=0\;.
\label{initial}
\end{equation}
In fact, our results are {\it not} 
sensitive to the choice of the initial data
provided that $|\phi(t_i)|,|\dot\phi(t_i)|\ll 1$, i.e., the scalar
field evolves from well below the Planck scale. 

At the first stage, the evolution of the Universe is essentially the
conventional Friedmann expansion. As $A'(0)=\delta>0$, the third term
in eq.~(\ref{Einstein-f}) acts as drag force, and positive scalar
field develops. The scalar field continues to increase as long as the
terms due to the scalar potential in eqs.~(\ref{Einstein-a}) and
(\ref{Einstein-f}) are negligible. At small $V(\phi)$ (e.g., small
$\lambda$ in eq.~(\ref{V1**})), the scalar field reaches its maximum
value $\phi_{\rm max}$ which depends on $\rho_0$ and $\delta$ (and, in
general, on the functional form of $A(\phi)$). When the scalar
potential starts to dominate the expansion, the  inflationary
stage begins. The scalar field slowly rolls down towards the origin
(provided that the Einstein-frame effective potential $U(\phi)$ slowly
increases as function of $\phi$ at $\phi\leq \phi_{\rm max}$), and the
Universe expands exponentially. Inflation ends at $\phi\sim\mu$ when
the scalar field begins to oscillate about $\phi=0$.  The number of
inflationary $e$-foldings depends on $\phi_{\rm max}$ and the
parameters entering $A(\phi)$ and $V(\phi)$.

The analysis of the first Friedmann stage is most easily performed
at $\rho_0\ll1$, i.e., when the Universe is curvature dominated 
right after the Planck epoch. Let us consider this case in some detail. 
We will see that $\phi_{\rm max}\sim\rho_0$, so one can approximate $A(\phi)$
by two first terms in eq.~(\ref{A=1+df}). Let us write
\[
a(t)=t+b(t)
\]
with $b\sim\rho_0$. Also, $\phi\sim\rho_0$. The scalar potential is
negligible at the first stage, and, to the linear order in $\rho_0$, we 
find from eqs.~(\ref{Einstein-a}) and (\ref{Einstein-f}),
\[
\frac{2}{t^2}\dot{b}+\frac{\delta}{t}\dot{\phi}=\frac{\rho_0}{t^\alpha}\;,
\]
\[
\ddot{\phi}+\frac{3}{t}\dot{\phi}-\delta\left(\frac{1}{t}\ddot{b}+
\frac{2}{t^2}\dot{b}\right)=0\;.
\]
The solution to these equations with the initial data~(\ref{initial}) 
imposed at $t_i=1$ (i.e., when $a(t_i)=1$) is
\begin{equation}
\phi=\phi_{\rm max}-\frac{\delta}{1+\delta^2/2}
\frac{\rho_0}{2}\left[\frac{1}{(\alpha-2)t^{\alpha-2}}-\frac{1}{2t^2}
\right]\;,
\label{D3*}
\end{equation}
where
\begin{equation}
\phi_{\rm max}=\frac{\delta}{1+\delta^2/2}
\frac{\rho_0}{2}\left[\frac{1}{\alpha-2}-\frac{1}{2}
\right]\;.
\label{D3**}
\end{equation}
The solution (\ref{D3*}) rapidly tends to $\phi_{\rm max}$ at $t\gg1$
(recall that $\alpha>3$). Hence, the initial data for inflation are
independent of the scalar potential at small $V(\phi_{\rm max})$. Note
that at least at small $\rho_0$, the maximum value of the scalar field
is small in Planck units, $\phi_{\max} \ll1$. 

The second, inflationary stage begins when $a(t)$ becomes of the order
of $[V(\phi_{\max})]^{-1/2}$ which is much greater than 1. In other
words, the duration of the first Friedmann stage is fairly large.

Let us now discuss the inflationary stage. At this stage
both the matter and spatial curvature terms in eq.~(\ref{Einstein-a})
may be neglected. At small enough $\mu$, the maximum value of the
scalar field, eq.~(\ref{D3**}), is such that eq.~(\ref{V=A2}) holds. Let us
write at $\phi\gg\mu$
\[
V(\phi)=\lambda A^2(\phi)\left[1+w(\phi)\right]\;,
\]
where $\lambda$ is a positive small constant, and $w(\phi)\ll1$ at 
$\mu\ll\phi\lsim\phi_{\max}$. The conditions of slow roll in this model are
\begin{equation}
\dot{\phi}\ll HA,\ \ \ \dot{\phi}\ll HA',\ \ \ \dot{\phi}\ll 
H\frac{A}{A'}, \ \ \ \ddot{\phi}\ll H\dot{\phi}\;.
\label{D4*}
\end{equation}
Under these conditions, the Hubble parameter has the form
\[
H=\sqrt{\frac{V}{A}}\left(1+h\right)\;,
\]
where $h\ll 1$. We find from eq.~(\ref{Einstein-a})
\[
h=-\frac{1}{2\sqrt{\lambda}}\frac{A'}{A^{3/2}}\dot{\phi}\;,
\]
which is indeed small. Equation~(\ref{Einstein-f}) then gives
\[
\dot{\phi}=-\displaystyle\frac{1}{3}\displaystyle\sqrt{\lambda}
\displaystyle\frac{A^{3/2}w'}{1+
A'^2/2A}\;.
\]
We find that the conditions~(\ref{D4*}) are indeed satisfied at small
$w(\phi)$, i.e., the Universe undergoes inflation in the slow roll regime.

Let us note in passing that the scalar field indeed rolls down towards
$\phi=0$ only if $w'(\phi)>0$. The opposite case corresponds to run-away
behavior without end of inflation. Generally speaking, the undesirable
run-away solutions appear when $V'(\phi)<2V(\phi)A'(\phi)/A(\phi)$,
i.e., when the Einstein-frame effective potential~(\ref{effective})
decreases at large enough $\phi$. In this paper we consider the
physically interesting case $V'>2VA'/A$, i.e., $w'>0$.

Finally, the number of inflationary $e$-foldings is
\begin{eqnarray}
N_e&=&\int_{\mu}^{\phi_{\max}}\frac{H}{|\dot{\phi}|}\,d\phi\nonumber\\
&=&\int_{\mu}^{\phi_{\max}}\frac{3\left(1+
A'^2/2A\right)}{Aw'}\, d\phi\;. \nonumber
\end{eqnarray}
This number is large at small $w(\phi)$ and depends both on the
properties of the Universe at the initial Friedmann stage (through
$\phi_{\max}$ ) and on the parameters of the model. As an example, for the 
concrete choice~(\ref{V1**}) one has at small $\rho_0$
\[
N_e=\frac{3}{8}\left(1+\frac{1}{2}\delta^2\right)
\frac{\phi_{\max}^4}{\mu^2}\;,
\]
where $\phi_{\max}$ is given by eq.~(\ref{D3**}). To get an idea of
numerics, at $\rho_0=1,\ \ \delta=1$, and $\gamma=0$ (non-relativistic
matter at the initial stage) one finds in
this example that $N_e\gsim 60$ at $\mu\lsim2\cdot 10^{-3}$. We see
that our scenario does not necessarily lead to an enormous number of
$e$-foldings.

Until now we have considered the homogeneous and isotropic Universe
filled with a specific form of matter. We argue, however, that the
non-minimal scalar field with the the property (\ref{V=A2}) is capable
of producing inflation from quite general initial state of the
Universe (the exceptions being purely radiation dominated FRW Universe
with zero scalar curvature and closed Universe of small size). In
general, the Universe at the initial stage has non-vanishing scalar
curvature.  Even if $R$ is relatively small at this stage, it drags
the scalar field out to the plateau of Fig. 1, albeit inhomogeneously,
provided that $\mu$ is small enough\footnote{In a sense, the above
case of the open Universe and small $\rho_0$ is particularly
unfavourable for establishing the initial conditions for inflation:
the Hubble parameter determining the friction term in
eq.(\ref{Einstein-f}) is large whereas the drag force is small
(proportional to $\rho_0$) and rapidly decreases in time.}. As the
Einstein-frame effective potential $U(\phi)$ is essentially
independent of $\phi$ at this plateau, the inflationary stage is
generic, and possible large inhomogeneities are stretched out.

To conclude, inflation may emerge as an intermediate stage of the
evolution of the Universe under mild assumptions on the cosmological
initial conditions. This feature, however, is inherent only in a
restricted class of models of scalar fields interacting with gravity:
the most essential ingredient of our scenario is the relation
(\ref{V=A2}) between the scalar potential $V(\phi)$ and
inflaton--gravity coupling $A(\phi)$.  It remains to be understood
whether this property of the scalar field may be motivated by models
of particle physics.

\paragraph{Acknowledgments.}

The work is supported in part by Award No. RP1-187 of the
U.S. Civilian Research \& Development Foundation for the Independent
States of the Former Soviet Union (CRDF), and by Russian Foundation
for Basic Research, grants 96-02-17804a and 96-02-17449a.


\begin{thebibliography}{99}
\bibitem{reviews} A.D.~Linde, {\it Particle Physics and Inflationary
Cosmology} (Harwood, Chur, Switzerland, 1990); E.W.~Kolb, M.~Turner,
{\it The Early Universe} (Addison-Wesley, New York, 1990); 
A.D.~Linde,
{\it Lectures on Inflationary Cosmology}, Stanford University preprint 
SU-ITP-94-36, hep-th/9410082. 
D.H.~Lyth, A.~Riotto,{\it Particle Physics
Models of Inflation and the Cosmological Density Perturbation},
Lancaster University  preprint LANCS-TH-9720,
hep-ph/9807278 (1998).
\bibitem{Guth} A.H.~Guth, Phys. Rev. {\bf D23}, 347 (1981).
\bibitem{new} A.D.~Linde, Phys. Lett. {\bf B108}, 389 (1982);
A.~Albrecht, P.J.~Steinhardt, Phys. Rev. Lett. {\bf 48}, 1220 (1982).
\bibitem{Linde-chaotic} A.D.~Linde, Phys Lett. {\bf B129}, 177 (1983).
\bibitem{Kaiser} 
N.~Makino, M.~Sasaki, Prog. Theor. Phys. {\bf 86}, 103 (1991);
D.~Kaiser, Phys. Rev. {\bf  D52}, 4295 (1995).
\end{thebibliography}
\end{document}